\documentstyle[12pt]{article}
\begin{document}
\begin{center}
\Large \bf Consequences of a covariant Description of
Heavy Ion Reactions at intermediate Energies\footnote{Supported
by GSI-Darmstadt and BMFT under contract 06T\"U736.} \\ \vspace*{5mm}
\normalsize
\quad\\ \vspace*{9mm}
\large E. Lehmann, Rajeev K. Puri, Amand Faessler, \\ G. Batko and
S.W. Huang
\\ \normalsize
\vspace*{6mm}
Institut f\"ur Theoretische Physik, Universit\"at T\"ubingen,\\
Auf der Morgenstelle 14, 72706 T\"ubingen, Germany
\\ \vspace*{5mm}
\today \\ \vspace*{5mm}
\end{center}
\vspace*{12mm}
%%%%%%%%%%%%%%%%%%%%%%%%%%%%%%%%%%%%%%%%%%%%%%%%%%%%%%%%%%%%%%%%%%%%%%%%%%%
\it Abstract: \\ \\
\normalsize
\baselineskip 24pt
Heavy ion collisions at intermediate energies are studied by
using a new RQMD code, which is a covariant generalization of
the QMD approach.
We show that this new implementation is able to produce the same results
in the nonrelativistic limit (i.e. 50MeV/nucl.) as the non-covariant QMD.
Such a comparison is not available in the literature.
At higher energies (i.e. 1.5 GeV/nucl. and 2 GeV/nucl.) RQMD and QMD give
different results in respect to the time evolution of the phase space,
for example for the directed transverse flow.
These differences show that consequences of a covariant
description of heavy ion reactions within the framework of RQMD are
existing even at intermediate energies.
\\ \\ \\
\newpage
%%%%%%%%%%%%%%%%%%%%%%%%%%%%%%%%%%%%%%%%%%%%%%%%%%%%%%%%%%%%%%%%%%%%%%%%%%
\begin{center}
\large \bf 1. Introduction
\end{center} \normalsize \medskip
\baselineskip 24pt
The only way to probe nuclear matter under extreme conditions of density
and thermal excitations is
to study heavy ion reactions.
The main aim of these experiments at intermediate energies, i.e.
between 100 MeV/nucl. to about a few GeV/nucl., is to get
some information on the nuclear equation of state (EOS),
which is characteristic of the physical structure of the
considered system. A knowledge of the EOS is of immense
theoretical and practical importance for nuclear physics as well
as astrophysics.

In the last decade strong efforts were directed to develop microscopic
models to describe the dynamics of heavy ion collisions at intermediate
energies.
Today mainly two different semiclassical theoretical approaches are used
with big success:\\
One of them, the so called BUU-type models
\cite{Bonasera89,Bertsch88,Gregoire87,Bauer86,
Kruse85,Bertsch84} simulate with help
of the test particle method kinetic equations like the
Boltzmann-\"Uhling-Uhlenbeck equation (BUU-equation)
which describes the time evolution of the one-body distribution function
in phase space.
The other one, the so-called "Quantum" Molecular Dynamics (QMD)
\cite{Aichelin91,Aichelin86} is based
on the classical Molecular Dynamics and hence propagates the particles under
mutual interactions. In addition also some important quantum theoretical
characteristics, which are also contained in the BUU-type models, like
Fermi motion of the nucleons, stochastic
scattering and Pauli blocking are included.

In QMD and the BUU-type approaches the nucleon-nucleon interaction
is splitted into a
long-range and a short-range part. While the long-range part influences
the particle trajectories in a steady way the short-range part is
responsible for the so called "hard collisions'', in which strong
changes of the momenta of the particles can appear.

With the help of such phase-space simulations a more clear picture of the
dynamics of heavy ion reactions has been created. But the main aim of
studying heavy ion collisions at intermediate energies, namely to
extract some information on the EOS,
could not be reached due to the sensitivity of the observables,
like the collective flow, to different ingredients.

One of these questions that still remains is how strong do
relativistic effects influence the dynamical evolution of a heavy ion
reaction at intermediate energies. Since even at this energy
range the nucleons are moving with a velocity which is not
negligible with respect to
the speed of light one has to work in a covariant framework to get
a reasonable description of heavy ion collisions at intermediate energies.
Therefore, it was worth to develop covariant generalizations of these
non-covariant microscopic models listed above.

Unlike the covariant generalizations of the BUU-type models
\cite{Ko87,Cassing90b,Fuchs92}, which
are closely connected to field theoretical features,
the covariant generalization
of the QMD approach is not straight forward. The reason for these difficulties
is that the mutual nucleon-nucleon interaction in QMD is dealt as a
simultaneous action-at-a-distance. Hence one enters the problems
of a
Poincar\'e invariant action-at-a-distance if one wants to
generalize the QMD approach in a manifestly covariant way.

But these problems can be treated in the framework of Constrained
Hamilton Dynamics. This formalism was introduced by Dirac \cite{Dirac49}
to express a
theory based on a singular Lagrangian in a generalized phase space approach
with help of constraints and later on extended by others
\cite{Todorov76,Rohrlich79,Komar78,Sudarshan81,
Samuel82} to a form
which is suitable for our purpose.
These authors picked up the
idea of Dirac \cite{Dirac49} who has realized for the first time that
constraints are not only reducing the degrees of freedom but can
also determine the dynamics.

Of course, all these covariant models are based on analytical expressions
with a well defined nonrelativistic limit. However, for none of
them it was shown that the numerical procedures used in the implementations
give in the nonrelativistic limit really the same results as the
corresponding non-covariant model. But one should make sure that
the covariant codes are working well at nonrelativistic energies
by checking sensitive quantities like the transverse flow before
one is able to discuss possible relativistic effects at higher energies.
This was also not done in ref.\cite{Maruyama91a} because these authors studied
only insensitive quantities at nonrelativistic energies.

Therefore we present here a detailed analysis of phase space
simulations at 50 MeV/nucl.. These results show that our
implementation of RQMD gives the same results as the
non-covariant QMD in the nonrelativistic limit.
Another aim of this paper is to study if this covariant description
of heavy ion reactions in the framework of RQMD has consequences
for the time evolution of the phase space in the intermediate
energy range. This is done by comparing results of sensitive
quantities like the directed transverse flow extracted from QMD
and RQMD calculations.

The paper is organized as follows:\\
In the following section we will briefly discuss the formalisms used in
QMD and RQMD and will point out their main differences.
The third section contains a few details of the implementation
of these methods.
In the forth section results of simulations of semicentral Ca+Ca
collisions at different energies are presented. These results
are analyzed in respect to discuss the differences of the used
approaches, e.g. QMD and RQMD.
Finally, in the last section, we summarize and also give an outlook.
\\
%%%%%%%%%%%%%%%%%%%%%%%%%%%%%%%%%%%%%%%%%%%%%%%%%%%%%%%%%%%%%%%%%%%%%
\newpage
\begin{center}
\large \bf 2. The Formalism used in QMD and its covariant Generalization
\end{center} \normalsize \medskip
\baselineskip 24pt
In this section we give a brief description of both, the QMD
approach and its covariant generalization, the RQMD approach.
For more details, especially for the foundations of these
formalisms, we refer to \cite{Aichelin91} in case of QMD and to
\cite{Sorge89} in case
of RQMD.

First we concentrate on the\\
\bf Quantum Molecular Dynamics (QMD): \normalsize \\ \baselineskip 24pt
The dynamics of heavy ion reactions can be studied in a
dynamical many body approach on an event by event base using QMD.
This model gives a microscopic description of heavy ion
collisions at the nucleon level and was developed by Aichelin
and St\"ocker \cite{Aichelin86}.

Each nucleon is represented by its Wigner density in phase space
in a Gaussian parametrization
\begin{equation}
f_i(\vec r,\vec p,t) = \frac{1}{\pi^3}\mbox{exp}[-(\vec p - \vec p_i(t))
^22L - (\vec r - \vec r_i(t))^2 / 2L] .
\end{equation}
with a fixed width.
The initial distribution of the two nuclei in orbital-space and momentum-space
are generated by a standard Monte-Carlo procedure by taking care
of the right radii of the nuclei, the Fermi energy and an
acceptable binding energy of each nucleus.
In order to simulate
heavy ion collisions this two well prepared nuclei are boosted towards each
other.

Since the nucleon-nucleon interaction in QMD is splited into a
long range part and a short range part, we calculate the time
evolution of the system of the two colliding nuclei in two parts:\\
1.) As the width of the Gaussians is fixed the centroids of each
nucleon $\vec r_i(t)$, $\vec p_i(t)$ are propagated by the
classical equations of motion
\begin{equation}
\frac{d\vec r_i}{dt} = \frac{\partial H}{\partial \vec p_i}
\end{equation}
\begin{equation}
\frac{d\vec p_i}{dt} = -\frac{\partial H}{\partial \vec r_i} ,
\end{equation}
where the Hamiltonian is given by the classical $N$-body Hamiltonian
\begin{equation}
H=\sum_i\frac{\vec p_i^{\,2}}{2m_i} + \frac{1}{2}\sum_{i,j}U_{ij}^{(2)}
+\frac{1}{3!}\sum_{i,j,k}U_{ijk}^{(3)} .
\end{equation}
The potentials in equation (4) are calculated as classsical exspectation
values by folding the two-
and three-body parts of the interaction with the Wigner
densities of the interacting nucleons:
\begin{equation}
U_{ij}^{(2)} = \int f_i(\vec p_i,\vec r_i,t)f_j(\vec p_j,\vec r_j,t)
               V_I^{(2)}(\vec r_i,\vec r_j,\vec p_i,\vec
p_j)d^3\vec r_i d^3\vec r_j d^3\vec p_i d^3\vec p_j
\end{equation}
\begin{equation}
U_{ijk}^{(3)} = \int f_i(\vec p_i,\vec r_i,t)f_j(\vec p_j,\vec r_j,t)
           f_k(\vec p_k,\vec r_k,t) V_I^{(3)}(\vec r_i,\vec r_j,\vec r_k)
d^3\vec r_i d^3\vec r_j d^3\vec r_k d^3\vec p_i d^3\vec p_j
d^3\vec p_k .
\end{equation}
Shrinking on  a local Skyrme force only one gets for the
total potential energy by using (5) and (6)
\begin{equation}
V_{Skyrme}=\sum_{i=1}^N\left[\frac{\alpha}{2}\left(\sum_{j\neq i}
           \frac{\tilde\rho_{ij}}{\rho_0}\right) +\frac{\beta}{\gamma+1}
           \left(
           \sum_{j\neq i}\frac{\tilde\rho_{ij}}{\rho_0}\right)^{\gamma}
           \,\right] ,
\end{equation}
with the so-called interaction density
\begin{equation}
\tilde\rho_{ij} = \frac{1}{(4\pi L)^{3/2}}\mbox{exp}(-\vec r_{ij}^{\,2}/4L) ,
\end{equation}
where $\vec r_{ij}$ is the distance of the centers of two Gaussians.

2.) If the centroids of two Gaussians come closer than a certain
distance $d_{min}=\sqrt{\sigma_{TOT}(\sqrt{s})/\pi}$ during their propagation,
a stochastic collision of the
two corresponding nucleons is calculated by a Monte-Carlo
procedure. In order to respect the Pauli principle the collision
will be blocked if the phase space elements of the final states
are already occupied by other nucleons. The collisions determined
in this way can be elastic or inelastic. The inelastic channels
included in the calculations presented in this paper are
creation and reabsorption processes of the $\Delta(1232)$ resonance.
The whole collision part is dealt by using a full relativistic
kinematic as explained in ref.\cite{Bertsch88}.

In the propagation part one can also include some relativistic
kinematic by replacing the kinetic energies in the Hamiltonian (4)
by $\sqrt{\vec p_i^{\,2}+m_i^2}$. This small modification is
always used in QMD calculations at relativistic energies.
But a manifestly covariant
description also requires a covariant formulation of the interaction.
Only in such a full covariant model one can make sure that
Lorentz scalar observables are independent of the observer frame.
How one can generalize QMD to a manifestly covariant model will
be described in the next subsection.\\ \\
\bf Relativistic Quantum Molecular Dynamics (RQMD):\normalsize \\ \\
\baselineskip 24pt
The QMD approach contains as an essential part the classical propagation
of particles under instantaneous mutual interactions ''at-a-distance''. The
generalization of this nonrelativistic particle dynamics to a
manifestly covariant particle dynamics is not trivial, because
one has to know how to deal action-at-a-distance in a covariant
manner. The conceptual problems in this field are formulated in
the famous no interaction theorem \cite{Currie63}, which states that
a Hamiltonian description of a multi-body system with a
canonical representation of the Poincar\'e group, where
world line invariance is demanded and the
physical coordinates are identified with the canonical coordinates
is incompatible with interaction.
This negative implication of the no interaction theorem can be avoided
by dealing not with the whole phase space but with a sub-manifold of it.
One possibility in this fashion is
given in the framework of Constrained Hamilton Dynamics.

In a covariant theory one has to respect Poincar\'e invariance and
hence every particle has to be described by its four momentum $p_i^{\mu}$
and its four position vector $q_i^{\mu}$, means every particle
carries its own time coordinate. Therefore, a
$N$-particle system is connected with an $8N$ dimensional phase space.
In the formalism of Constrained Hamilton Dynamics
this phase space is reduced to an $6N$ dimensional phase
space with the help of $2N$ constraints fixing the energies and the
relative times of the particles. In addition a global evolution
parameter has to be introduced by these constraints in order to
gauge the evolution of the system. Doing this
one defines a $6N$ dimensional hypersurface in the original $8N$
dimensional phase space on which the system is allowed to move
during its evolution.
This formalism is not changing the
notion of simultaneity by a change of the frame of reference,
means one gets an invariant notion of simultaneity
as well as invariant world lines in this fashion.

The QMD model was extended to its covariant version, the RQMD model,
first by Sorge et al. \cite{Sorge89}
to study heavy ion collisions at ultrarelativistic energies. We
use a similar method in order to study heavy ion collisions at the
intermediate energy range. In the following we will discuss some
details of this method.

We use here similar constraints as introduced in ref.\cite{Sorge89}
and hence the first $N$ constraints are chosen as on-shell
constraints\footnote{We use the Einstein convention for the tensor
indices, but no summation over repeated particle indices, except
if explicitly specified.}
\begin{equation}
K_i = p_i^{\mu}p_{i\mu} - m_i^2 -\tilde V_i =0 \qquad ; \qquad i=1,...,N
,\end{equation}
which request that the particles move between collisions on
energy shell. This choice of the first $N$ constraints require that
the potential part $\tilde V_i$ should be a Lorentz scalar and therefore
a function of Lorentz scalars. Since we want to define a
system with mutual two-body interactions like in QMD, $\tilde
V_i$ should be given by a sum of these two-body interactions.
Following ref.\cite{Sorge89} we use therefore
\begin{equation}
\tilde V_i = \sum_{j\neq i}\tilde V_{ij}(q^2_{Tij})
,\end{equation}
which means that the two-body interactions depend only on the
Lorentz invariant squared transverse distance
\begin{equation}
q^2_{Tij} = q^2_{ij} - \frac{(q^{\mu}_{ij}p_{ij\mu})^2}{p_{ij}^{\,\,2}}
,\end{equation}
with $q^{\mu}_{ij}=q^{\mu}_i-q^{\mu}_j$ being the four
dimensional distance and $p^{\mu}_{ij}=p^{\mu}_i+p^{\mu}_j$ the
sum of the momenta of the two interacting particles $i$ and $j$.

Motivated by a comparison in the non-relativistic limit we use for
the potential part finally
\begin{equation}
\tilde V_i =  2m_i\,\left[
\frac{\alpha}{2} \left(
\sum_{j\neq i}\frac{\mbox{exp}[q_{Tij}^2/4L]}{\rho_0(4\pi L)^{3/2}}
\right)
+\frac{\beta}{(\gamma+1)} \left(
\sum_{j\neq i} \frac{\mbox{exp}[q_{Tij}^2/4L]}{\rho_0(4\pi L)^{3/2}}
\right)^{\gamma} \,\right] .
\end{equation}

In this way the local Skyrme interaction used in QMD is
generalized by replacing the squared distance $-\vec r^{\,2}_{ij}$ in the
interaction densities (8) by the Lorentz invariant squared transverse
distance $q^2_{Tij}$. One should notice that due to the second
term of the right
hand side of equation (11) this generalized interaction
used in RQMD is slightly implicit momentum dependent. Because
of this term, which gives the longitudinal squared distance, the
interaction used in RQMD depends not only on the distance of the two
interacting particles, as in QMD, but also on the direction of
their center of mass motion in the rest frame of the two nuclei.

Since the on-shell constraints alone do not specify the world
lines one needs additional $N$ constraints which are fixing the
relative times of the particles.
In order to respect world line invariance, causality and cluster
seperability this $N$ time fixations are defined as
\begin{equation}
\chi_{i} = \sum_{j(\neq i)}g_{ij}p^{\mu}_{ij}q_{ij\mu} = 0 \qquad ;
\qquad i=1,...,N-1
\end{equation}
\begin{equation}
\chi_{N} = P^{\mu}Q_{\mu}-\tau = 0
\end{equation}
with $P^{\mu}=p^{\mu}/\sqrt{p^2}$, $p^{\mu}=\sum_i p^{\mu}_i$,
$Q^{\mu}=\frac{1}{N}\sum_i q^{\mu}_i$ and the dimensionless scalar
weighting function
\begin{equation}
g_{ij} = \frac{1}{q^2_{ij}/L_C}\mbox{exp}(q^2_{ij}/L_C)
\end{equation}
with $L_C=8.66$ fm$^2$. The conditions (13) are motivated by
studies in the framework of Singular Lagrangians \cite{Dominici78}.
Using this methods one gets up to the weighting functions
$g_{ij}$ the same conditions as secondary constraints and the
on-shell conditions (9) as primary constraints in a natural way.
The important fact for using the expression (15) as weighting
function is that this scalar function respects the principle of causality,
while the ones used in the Singular Lagrangian theories and in
the model of Samuel \cite{Samuel82} can violate
this important physical restriction.

The constraints (13) take care that the times of interacting particles
are not dispersed too much in their common CMS$_{ij}$.
Furthermore, they specify the dynamics by fixing the times at
which the forces has to be calculated but they do not specify the
global evolution parameter $\tau$. This evolution parameter must
defined to be determined dynamically since the no-interaction theorem
can only be avoided in this way as pointed out in ref.\cite{Sudarshan81}.
The evolution parameter $\tau$ is defined by the gauging condition
(14) in a way, that the individual times are
increasing with increasing $\tau$.

This set of $2N$ constraints given by (9),(13) and (14)
determines covariant world lines parametrized by the initial
data at equal $\tau$, means the starting conditions
are given by the values of $p^{\mu}_i$ and $q^{\mu}_i$ at a
given starting value of $\tau$.

With the help of this set of constraints the reduction of the
phase space can be done in a well defined way and the dynamics of
the system can be determined by the Hamiltonian
\begin{equation}
H =  \sum_{i=1}^{N}\lambda_i\,K_i
    + \sum_{i=1}^{N-1}\lambda_{N+i}\,\chi_i
,\end{equation}
given by a linear combination of the
$\tau$ independent $2N-1$ constraints
and hence, in the sense of Dirac \cite{Dirac49}, these constraints determine
the dynamics. The Hamiltonian (16) generates equations of motion
with the help of Poisson brackets as
\begin{equation}
\frac{dq_i^{\mu}}{d\tau} = [H,q_i^{\mu}]
\end{equation}
\begin{equation}
\frac{dp_i^{\mu}}{d\tau} = [H,p_i^{\mu}]
.\end{equation}

The unknown Lagrange multipliers $\lambda_i$ can be determined using the
fact that the complete set of $2N$ constraints must be fulfilled during
the whole time evolution.

This formalism has a well defined nonrelativistic limit as
shown in ref.\cite{Sorge89}.
This fact is a prior condition
that RQMD and QMD calculations should give the same results at
nonrelativistic energies if one uses in RQMD the same type of
interaction as in QMD but generalized in the way as described
above for the Skyrme interaction.

The binary collisions are dealt in RQMD in the same way as in
the non-covariant QMD by using Monte Carlo methods.
But in RQMD we use a full covariant kinematic to determine
these collisions whereas in QMD the kinematic is relativistic, but not
manifestly covariant.\\

%%%%%%%%%%%%%%%%%%%%%%%%%%%%%%%%%%%%%%%%%%%%%%%%%%%%%%%%%%%%%%%%%%%%%%%%%
\begin{center}
\large \bf 3. Details of the Implementation
\end{center} \medskip \baselineskip 24pt
Before we present some results of calculations in the next
section we will discuss in this section some
details of the numerical realization of QMD and RQMD.

First of all we would like to stress, that in contradiction to
earlier RQMD codes \cite{Sorge89,Maruyama91a} our new code is fully
integrated in a simulation package which contains the
non-covariant QMD approach with its different options and the covariant
RQMD approach as well. The whole package is called UNISCO
standing for {\bf UNI}fied {\bf S}imulation
{\bf CO}de\footnote{A full overview of this
simulation package, which on the non-relativistic side
is based on the latest version of the QMD of
Aichelin and coworkers, is given in \cite{Lehmann93}.}.
In this development special care is given to keep exactly the
same initial conditions in both approaches, QMD and RQMD. In
addition, this way of implementation assures us to use the same
parameters in QMD and RQMD calculations.
This is extremely important when one wants to look for
relativistic effects at intermediate energies and guarantees that
disagreements of results extracted from these two approaches
have a physical origin.

In QMD/RQMD heavy ion collisions are simulated with the help of independent
runs. In each run a single event is calculated. In the full
simulation the average of these independent events is taken in
order to get representative results.

In each event two carefully prepared nuclei are boosted against each other.
A single nucleus is build up with help of a Monte Carlo sampling
method.
This procedure used in QMD
is applied in RQMD as well. But in RQMD one has additionally to take care
of the constraints. A violation of these constraints right from
the starting would be conserved during the whole time evolution
and hence one would leave the base of the formalism in such a case.
Therefore one can only accept distributions in RQMD which are able to
fulfill the constraints and special care is taken in the code to fulfill
this condition.

By this procedure one gets nuclei which are stable for a much longer
time span than the usually considered reaction time. This was already
carefully proven for QMD in ref.\cite{Aichelin91}. For RQMD we present
here results in figure 1. As a measure of the stability of the nuclei we
have plotted the time evolution of the root mean square radius $R_{rms}$
for various single nuclei, which are prepared by the procedure discussed
above. For all three nuclei ($^{12}$C,$^{28}$Si,$^{40}$Ca) the time
evolution of the root mean square radii
of 10 different initializations boosted  to a certain energy is
plotted. As can be observed from figure 1, very light nuclei like
$^{12}$C suffer strong vibrations whereas the root mean square radii
of heavier nuclei, like $^{40}$Ca, are more smooth.
Finally, these results show clearly, that these nuclei, which are even not
heavy
ones, are stable for a very long time span (at least 200 fm/c).

In the part of the code where the centres of the Gaussians are propagated
we integrate the Hamilton equations (2)
and (3) in the case of QMD and the equations of motion generated
by the formulae (17) and (18) in the case of RQMD. The numerical
integrations are done by standard integration routines.

In order to decide if a baryon-baryon collision will occur or
not we use a strictly geometrical interpretation of the cross
section. Therefore two baryons will collide if their distance
become closer than $d_{min}=\sqrt{\sigma_{TOT}(\sqrt{s})/\pi}$
within a small time interval, whereas those passing each other
at a larger distance will not suffer a collision during this
time interval. $\sigma_{TOT}(\sqrt{s})$ is the total
nucleon-nucleon cross section at a given c.m. energy of the
colliding nucleon system. The time interval is given by the
actual time step size of the numerical integration used in the
propagation part.

Using this minimal distance concept in QMD one compares $|\vec
r_i-\vec r_j|$ with $d_{min}$ in order to decide if the
particles $i$ and $j$ became canditates for a collision. Since
this collision criterion is not Lorentz invariant the collision
sequence and the number of collisions can depend on the frame of
reference as has been shown in ref. \cite{Kodama84}. In RQMD we
use an invariant measure, namely the Lorentz invariant
transverse distance by comparing $\sqrt{-q_{Tij}^2}$ with $d_{min}$
and we are therefore avoiding these difficulties.

The scattering angles of a single baryon-baryon collision are determined
randomly by a standard Monte Carlo procedure
whereas the magnitudes of the
final momenta are fixed by conservation of energy and momentum.
For the cross sections used in this procedure we use the so
called Cugnon parametrization \cite{Cugnon88} of the free nucleon-nucleon
scattering data, which includes excitation and reabsorption processes
of the $\Delta(1232)$ resonance and takes care of the different
isospin channels of the processes included.

Whenever the final state of a nucleon-nucleon collision is
determined in this way we compute for each scattered nucleon the overlap
in phase space with the surrounding nucleons. In this procedure
we assume that each nucleon occupies a sphere in coordinate and
momentum space. Therefore, in RQMD this overlap has to be
determined in the rest frame of the nucleon of interest to
justify the assumption of a spherical distribution of the nucleon.
In QMD the overlap is calculated in the CMS of the two nuclei and
hence a spherical distribution of the nucleon is not ensured in the
non-covariant approach. From this
overlap in phase space one can then determine the probability
with which the collision is blocked. Using this procedure one is
able to respect the Pauli principle on a semiclassical level in
a satisfying way as proven in ref.\cite{Aichelin91}.

%%%%%%%%%%%%%%%%%%%%%%%%%%%%%%%%%%%%%%%%%%%%%%%%%%%%%%%%%%%%%%%%%%%%%
\newpage
\begin{center}
\large \bf 4. Results and their Discussion
\end{center} \medskip\ \baselineskip 24pt
1. Nonrelativistic Limit:\\ \\
As we have already stressed in this paper, it makes absolutely
no sense to discuss possible relativistic effects without making
sure, that the methods used for this studies, in our case QMD
and RQMD, do not give the same results in the nonrelativistic
limit. Therefore we present here first a comparison of QMD and RQMD at
50 MeV/nucl., which is an energy where no relativistic effect
should occur.

As a measure of the evolution of the phase space we have extracted three
different quantities from QMD/RQMD calculations: the nucleon density in the
central zone of a heavy ion reaction as a measure how much the
nuclear matter was compressed during the reaction, the rapidity
distribution as a measure of the stopping and the directed
transverse momentum as a measure of the bounce off. The last
quantity is defined as
\begin{equation}
p_x^{dir} = \frac{1}{N}\sum_{i=1}^{N}\mbox{sign}[Y_i^{(CM)}]p_{ix}
\end{equation}
with the rapidity of the i-th particle evaluated in CMS from
\begin{equation}
Y_i = \frac{1}{2} \mbox{ln} \left(\frac{E_i+p_{iz}}{E_i-p_{iz}}\right).
\end{equation}
Another measure of the bounce off, namely the transverse momentum
distribution $p_x/A$ as a function of the rapidity, is often
used instead of $p_x^{dir}$, especially to compare calculations with
experimental data. In this paper we are interested in a relative
comparison of two theories and hence we are using here $p_x^{dir}$,
which has the following advantages for our purpose:\\
In contradiction to $p_x/A$ the quantity $p_x^{dir}$ integrates over all
rapidity bins and provides therefore one value at a given time
and allows to follow this quantity
easily as a function of time. Due to this fact $p_x^{dir}$ as a
function of time reflects also the creation of the bounce off
during the reaction.

Figure 2 presents the results for the nucleon density as a
function of time, calculated in a sphere with 2 fm radius around
the origin of the CMS, the rapidity distribution of the final
state (after 120 fm/c) and the directed transverse momentum as a function
of time as extracted from QMD and RQMD calculations of semicentral
C+C collisions at 50 MeV/nucl.. The impact parameter (b=1.3 fm)
is half the radius of the two colliding nuclei.

As can be recognized from these plots, all three quantities reflect
their typical behaviour at low energies, e.g. the nucleon density
is in all cases far less than 2$\rho_0$ because of the small
compression at this low energy, the rapidity distribution shows one
maximum due to the small bombarding energy and the directed flow
is negative.

All results of these three quantities for semicentral C+C
collisions presented in figure 2 show an excellent agreement of
QMD and RQMD. Even the directed transverse flow, which is a
highly sensitive quantity in respect to all ingredients (physics
and numerics), shows this excellent agreement during the whole
reaction time.

The analysis of semicentral collisions
of a higher mass system, namely Ca+Ca, gives also a good agreement of
QMD and RQMD as shown in figure 3. In this figure the same three
quantities as in figure 2 are displayed. (The rapidity distribution
is analyzed at 100 fm/c in this reaction.) The small differences in the
directed flow (few \%)
are understandable by taking into account that the Constrained
Dynamics used in RQMD needs more numerical efforts than the
simple Hamilton Dynamics used in QMD.

Therefore, in conclusion, we can point out that our new RQMD
gives in the nonrelativistic limit the same results as the
non-covariant QMD as it should be. The fulfilling of this critical
bench mark convinces that one can study with help of this code
relativistic effects in heavy ion collisions
comparing  RQMD and QMD at higher energies.
\\ \\
2. Relativistic Energies:\\ \\
In order to test if the covariant description of heavy ion
reactions as given in RQMD shows already consequences at intermediate
energies we compare here QMD and RQMD calculations at 1.5
GeV/nucl. and 2 GeV/nucl..
At these energies the nucleons are moving with a velocity of about
85\% and 92\%
of the speed of light, respectively, and hence possible
relativistic effects should become visible at this energies.

We concentrate here our investigations on semicentral Ca+Ca
collisions and analyze one of the most sensitive quantities,
namely the directed transverse momentum as defined by eq.(19).
The impact parameter is 2 fm, which
is about half the radius of the considered nuclei.

In figure 4 the directed transverse momentum is displayed as a
function of time as extracted from QMD and RQMD calculations of
semicentral Ca+Ca collisions at a bombarding energy of 1.5 GeV/nucl..
In figure 5 results of the same reaction, but at 2 GeV/nucl. are plotted.

Analyzing the time evolution of the directed flow created during
the reaction one observes an appreciable difference among the different
equations of state and, more important for our intention, among QMD and RQMD.

In the early time stage of the reaction the directed flow is negative
since the interaction between the nuclei is attractive. After
the two nuclei collide the flow becomes positive because a
repulsion is induced by the high density in the region where the
two nuclei overlap each other. From this region particles are bounced off
due to their interactions with the other particles.

This repulsion is stronger in the case of the hard EOS and hence
the transverse flow becomes larger in this case as for the soft EOS. In
RQMD the interaction gets less attractive in the early time
stage and hence starts earlier to become repulsive than in QMD
as all results plotted in figure 4 and figure 5 are showing.

Due to the covariant treatment of the interaction in RQMD the
strength of the attractive forces, dominating the early time
stage of the reaction, is smaller in RQMD than in QMD. This can
be understood by the fact that the forces will be modified in the
covariant treatment by correcting factors which convert the
non-relativistic force to the spatial part of the corresponding
Minkowski force.

Furthermore, the final values of the directed flow
in RQMD and QMD are very sensitive to the energies considered.

At 2 GeV/nucl. as shown in figure 5, the flow is enhanced in
RQMD in comparison to QMD in case of the hard and the soft EOS
as well. At 1.5 GeV/nucl. (see figure 4) the flow is decreased in RQMD in
comparison to QMD for the soft EOS but increased for the hard EOS.
Therefore, the comparison of the results at 1.5 GeV/nucl. and 2 GeV/nucl.
plotted in figure 4 and figure 5, respectively, shows, that the
covariant description affects the flow strongly but not in a
unique way.

Information about the causes of the differences in the
flow can be achieved by comparing QMD and RQMD calculations in a
Vlasov-mode (no collisions) as well as in a cascade-mode
(collisions only). Whereas calculations in the Vlasov-mode give
a measure of the influence of the
mutual potential interactions on the flow, calculations in the
cascade-mode measure the influence of two-body collisions on the
flow. Results obtained by these calculations for semicentral Ca+Ca
collisions at 2 GeV/nucl. are
displayed in figure 6. These results show an increase of the
flow in RQMD calculations in the Vlasov limit compared to
equivalent QMD calculations. RQMD creates due to the Lorentz
contraction a higher density and therefore already a positive
flow in the final state of the considered reaction whereas
in QMD the flow is negative. This fact reflects that in QMD
due to the lower density
mainly attractive interactions are created.

In the
cascade limit the flow is decreasing in RQMD compared with QMD.
This difference has its reason in the different treatment of the
two-body collisions in the two considered approaches: Although QMD contains
a relativistic kinematic RQMD uses a full manifest covariant
kinematic, which avoids problems of reference frame dependent collision
sequences. Furthermore, in RQMD due to relativistic contractions
more collisions are Pauli
blocked than in QMD, which also reflects the different treatment
of the two-body collisions in the two approaches.

The comparison of the Vlasov limit and the cascade limit
makes clear, that the mutual potential
interactions and the two-body collisions affect the flow
in opposite directions in RQMD and in QMD. Therefore,
it becomes understandable, that one can get an increased as well
as a decreased flow when using RQMD instead of QMD, as it was
shown for semicentral Ca+Ca collisions at 1.5 GeV/nucl. in
figure 4.

The interplay of these two contrary effects is studied again
in figure 7, which shows a decomposition of the flow for the
same reaction as considered in figure 5.
In full QMD/RQMD calculations both
effects are acting together because they
are both present, whereas in the Vlasov limit or in the cascade limit
either the collisions or the mutual potential interactions are switched
off, respectively.
Here, the contribution
created by the mutual potential interactions contained in the Hamiltonian
as well as the contribution
created by two-body collisions is calculated
separately as a function of time.
This decomposition is
extracted from full QMD/RQMD calculations and is done in the following way:\\
In each time step the momentum transfer created by the mutual potential
interactions
and the momentum transfer produced by two-body collisions is
analyzed separately by using eq.(19).
Therefore, this analysis gives two values in each time step and both
contributions can be easily followed during the whole reaction.
If one adds both contributions one gets the total directed
transverse flow. From this analysis one
realizes, that in QMD the flow is mainly created by two-body collisions
in the compression stage in the considered reaction whereas the
mutual potential interactions even decrease the flow because they create a
negative contribution. In RQMD the collisions as well as the
mutual potential interactions influence the flow in the same direction and
nearly with the same amount. Therefore, one
gets finally a higher sidewards flow in RQMD in the considered reaction.
The contribution of the flow created
by mutual potential interactions reflects the same effect as was already
discussed
in the Vlasov limit: The mutual potential interactions alone create no positive
flow in QMD whereas in RQMD one gets already a positive
flow in the Vlasov limit in the considered reaction (see figure 6).
The enhancement of the repulsion in RQMD
in comparison to QMD, which is responsible for this effect,
is mainly created by the Lorentz contraction
of the nuclei in the CMS (compare also figure 8).

Figure 6 as well as figure 7 show, that the mutual potential
interactions create more positive flow in a covariant treatment whereas
binary collisions create less flow in RQMD than in QMD.
Comparing covariant approaches
with corresponding non-covariant approaches Ko et. al \cite{Ko87} as well as
Schmidt et. al \cite{Schmidt89}\footnote{This approach based on
classical molecular dynamics was only Lorentz invariant up to
the order of $(v/c)^2$.} found, that the flow is strongly
increased if binary collisions are disregarded. The flow might
be reduced also in their approaches if binary collisions are
taken into account.

In order to gain more information about the differences in the
flow we have done two
additional calculations at 2 GeV/nucl. for the same reaction,
which are shown for the hard EOS in
figure 8, together with the results already plotted in figure 5.

In one calculation, assigned with QMD*, we have changed the
starting conditions in QMD by using Lorentz contracted distributions
of the nuclei in orbital space and Lorentz elongated distributions
in momentum space in the CMS of the two nuclei.
The phase space is always initialized in this way in
RQMD but not in the usual QMD calculations.
Since we use only static
Skyrme forces in our studies presented in this paper
the nuclei are stable for a longer time span than the
reaction time also when this special initialization of the nuclei
is applied in QMD.

Comparing the results plotted in figure 8 one realizes that
QMD calculations with modified initial conditions results in
even larger values for the transverse flow in the expansion stage than with
RQMD. Due to the Lorentz contraction of the nuclei the density
is increased. Therefore, the repulsive interaction gets stronger
and the flow is increased. This effect is partly counterbalanced in
RQMD by the covariant treatment of the
interaction\footnote{One should note that this
result is an improvement over earlier RQMD calculations
\cite{Maruyama91b}.}.
But in comparison
to normal QMD calculations RQMD increases the transverse flow
for the reaction shown in figure 8.

In another calculation, assigned with RQMD* in figure 8, we have
modified RQMD in order to study the importance of the multi-time
description in this approach.
Our intention in this modification was to force the particles to
have the same time coordinate in a defined frame of reference
during the whole reaction. This can be not done by defining
simple non-relativistic time fixations like
\begin{equation}
\chi_i = q_i^0 -q_N^0 = 0 \qquad ; \qquad i=1,..,N-1
\end{equation}
\begin{equation}
\chi_N = q_N^0 - \tau = 0,
\end{equation}
because these constraints are not Poincar\'e invariant and hence
would strongly violate the foundations of the underlying formalism
of Constrained Hamilton Dynamics.
Therefore, in order to violate the restrictions of the formalism
as less as possible, we have used more refined
constraints even in this modification of RQMD and have replaced the
$\tau$-independent $N-1$ time fixations (13) by
\begin{equation}
\chi_i = P^{\mu}\,(q_{i\mu}-Q_{\mu}) = 0 \qquad ; \qquad i=1,...,N-1,
\end{equation}
whereas the gauging condition (14) is kept the same.
These set of time constraints given by eq.(23) and eq.(14) stresses
the particles to have equal time coordinates in the CMS. Therefore,
one can use this modification only for calculations in the CMS
if one wants to have equal time coordinates of all particles.
In addition, one should remark, that these time constraints are
not really an alternative to the constraints defined in eqs.(13)
and (14), because they violate cluster separability and
causality (compare \cite{Balachandran82}).
But condition (23) fixes that all particles have the same time
coordinate in the CMS. This allows to study the importance of
the multi-time description in RQMD requested by Lorentz
invariance.

The results displayed in figure 8 show clearly, that the difference
among RQMD and QMD is drastically reduced if one disregards the
multi-time description and requests equal times according to eq.(23)
in the RQMD approach. However, this multi-time
description presents the price which one has to pay to have a full
covariant formalism.
Therefore, it is not surprising that individual times of the
different baryons in RQMD
play an important role for relativistic effects in the
transverse flow. The covariant treatment of the interaction in
this multi-time description enhances the flow in comparison to
non-covariant QMD calculations in the considered reaction as can
be seen from figure 8.

In order to gain a better overview it is helpful to summarize
here the different influences on the flow connected with the
covariant description in the framework of RQMD:\\
1. Different starting conditions affect the flow to a certain amount.
A Lorentz contracted distribution in coordinate space enhances
the flow drastically. This effect is partly counterbalanced by
the covariant treatment of the interaction in RQMD. This
covariant treatment of the interaction can only be done in a
multi-time description, which takes care of individual time coordinates
of all particles. A disregard of this fact, which violates the foundations
of the underlying formalism, produces a flow
which is not much different to the one obtained by non-covariant
QMD calculations.\\
2. The ''long range part'' of the nucleon-nucleon interaction,
contained as mutual potential interactions in the Hamiltonian in the approaches
discussed,
increases the flow in RQMD in comparison to QMD, mainly due to
the Lorentz contraction of the nuclei. In addition, one should note,
that the implemented forces are slightly different
although they are based on the same static Skyrme
force in both approaches. This difference in the forces used is also
responsible for some differences of the flow between QMD and RQMD.\\
3. The directed sidewards flow is also affected by the covariant treatment
of the two-body collisions, which represent mainly
the ''short range part'' of the
nucleon-nucleon interaction. This
effects reduce the flow in RQMD compared with QMD, where
the treatment of the two-body collisions is only relativistic
but not manifestly covariant.

Finally, all these effects produce the
difference in flow between RQMD and QMD at relativistic energies
reflecting kinematical and dynamical relativistic effects.
The results for the directed transverse
flow show clearly, that a Lorentz invariant
treatment is necessary to study heavy ion reactions in the GeV
energy region.

%%%%%%%%%%%%%%%%%%%%%%%%%%%%%%%%%%%%%%%%%%%%%%%%%%%%%%%%%%%%%%%%%%%%%
\vspace*{1cm}
\begin{center}
\large \bf 5. Summary and Outlook
\end{center} \medskip \baselineskip 24pt
Heavy ion collisions were studied in the framework of QMD and RQMD
at relativistic energies as well as in the nonrelativistic limit.
The formalisms used in QMD and RQMD were discussed by working
out their main differences.

Although the nucleon-nucleon interaction is described in both
approaches by static Skyrme forces it turned out, that the
correct covariant treatment of this interaction, as done in RQMD,
results in some differences in comparison to the non-covariant
approach used in QMD. Further differences can be produced by a
manifest covariant treatment of the two-body collisions which
assures reference frame independent collision sequences.

It was proven explicitly, that our new implementation of the RQMD
approach produces the same results as QMD in the nonrelativistic
limit by comparing RQMD and QMD calculations of semicentral
C+C and Ca+Ca collisions at 50 MeV/nucl.. This critical bench mark is
even fulfilled for highly sensitive quantities like the directed
transverse flow during the whole reaction.

At relativistic energies the covariant description of heavy ion
reactions as given in the framework of RQMD shows consequences
on the dynamics of heavy ion reactions. These consequences were
studied by comparing the directed flow produced during semicentral
Ca+Ca collisions at 1.5 GeV/nucl. and 2 GeV/nucl. as extracted
from QMD and RQMD calculations.

The careful analysis of the observed differences turned out that
the flow is affected by several reasons, based on various
differences among QMD and RQMD.

A simple modification of the initial conditions in QMD towards
relativity, e.g. Lorentz contracted distributions in coordinate
space and Lorentz elongated distributions in momentum space, but
treating the interaction not covariant leads to a drastic
overestimation of the flow.
A disregard of the multi-time description, which presents the price
to pay for a manifest covariant treatment,
leads to a flow which is not much different in comparison to the
flow obtained by QMD calculations.
But the correct treatment of the dynamics in a multi-time
formalism produces quite different results for the flow in RQMD
compared to the flow in QMD.

Whereas the difference in the treatment of the mutual potential interaction
contained
in the Hamiltonian increases
the flow in RQMD
the manifest covariant treatment of the two-body collisions decreases
the flow in RQMD in comparison to QMD.
Finally, the interplay of these contrary effects is responsible for the
differences of the directed flow detected from comparisons of QMD
and RQMD calculations at relativistic energies.
Due to the fact, that these two effects act in opposite
directions, one can get an increased but also an decreased flow
in RQMD in comparison to QMD.

All RQMD calculations presented were obtained by describing the
mutual two-body potential interactions with help of generalized static
Skyrme forces, treated as scalar potentials in
Constrained Hamilton Dynamics. However, a more reasonable and
realistic description of heavy ion collisions should take care of
the complete Lorentz structure of the nucleon-nucleon interaction
and has to include large scalar and vector potentials as well.

Therefore, in future, on should work with modified on-shell
constraints given by
\begin{equation}
\tilde K_i = \Pi^{\mu}_i\,\Pi_{i\mu} - m_i^{*2} = 0,
\end{equation}
where the effective momenta
\begin{equation}
\Pi_i^{\mu} = p_i^{\mu} - g_v\,A_i^{\mu}
\end{equation}
contain all mutual two-body vector interactions, e.g.
$A_i^{\mu}=\sum_j\,A_{ij}^{\mu}$ and the effective masses
\begin{equation}
m_i^* = m_i + g_s\,\Phi_i
\end{equation}
contain all mutual two-body scalar interactions, e.g.
$\Phi_i=\sum_j\,\Phi_{ij}$.

In order to take care of the nuclear medium and the correct
energy and momentum dependence of the nucleon-nucleon
interaction one should work with realistic forces and hence,
these interactions should be extracted from self-consistent
Dirac-Br\"uckner calculations.

However, such a treatment is not at all trivial, but nevertheless,
work on this line is in progress.\\
\vspace*{2cm}\\
The authors are highly thankful to J. Aichelin, who kindly provided
us the new version of the QMD code which together with our RQMD code
is contained in the program package UNISCO. Also, two of us (EL and RKP)
acknowledge enlightning discussions with C. Fuchs, C. Hartnack,
T. Maruyama and H. Sorge.

%%%%%%%%%%%%%%%%%%%%%%%%%%%%%%%%%%%%%%%%%%%%%%%%%%%%%%%%%%%%%%%%%%%%%

%%%%%%%%%%%%%%%%%%%%%%%%%%%%%%%%%%%%%%%%%%%%%%%%%%%%%%%%%%%%%%%%%%%%%%%%
\newpage
\large \bf Figure Captions
\normalsize \baselineskip 24pt~\\
\vspace*{4mm}\\
{\bf Fig. 1:} Root mean square radii of various nuclei (C,Si,Ca)
as a function of time.\\ \\
{\bf Fig. 2:} Nucleon density (a) and directed transverse flow (c)
as function of time and the rapidity distribution (b) at the final
state (120 fm/c) of semicentral C+C collisions at 50 MeV/nucl.
as extracted from QMD and RQMD calculations. The impact
parameter was 1.3 fm and a static hard (left) as well as a
static soft (right) Skyrme interaction was used.\\ \\
{\bf Fig. 3:} Same as figure 2 but for semicentral Ca+Ca collisions.
The impact parameter was 2 fm and a static hard Skyrme
interaction was used. The rapidity distribution (b) is plotted
at the final state (after 100 fm/c).\\ \\
{\bf Fig. 4:} Comparison of QMD and RQMD calculations of
semicentral Ca+Ca collisions at a bombarding energy of 1.5 GeV/nucl.
and an impact parameter of 2 fm. The directed transverse flow is
plotted as a function of time. Results obtained with a hard EOS (above)
are shown as well as results obtained with a soft EOS (below).
\\ \\
{\bf Fig. 5:} Same as figure 4 but at a bombarding energy of 2
GeV/nucl..\\ \\
{\bf Fig. 6:} Directed transverse flow as a function of time as
obtained from QMD and RQMD calculations in a Vlasov-mode (above)
and a cascade-mode (below). Ca+Ca collisions at 2 GeV/nucl. and an
impact parameter of 2 fm were calculated by using a hard EOS.\\ \\
{\bf Fig. 7:} Different contributions to the directed transverse
flow as extracted from QMD and RQMD calculations for the same reaction
(Ca+Ca at 2 GeV/nucl., b=2 fm) as in figure 5.
Here the contribution of the directed
transverse flow created by mutual potential interactions (MPI)
contained in the Hamiltonian
and the contribution created by two-body collisions (Coll)
are displayed separately.\\ \\
{\bf Fig. 8:} Directed transverse flow as a function of time as
extracted from various QMD and RQMD calculations for semicentral
Ca+Ca collisions at a bombarding energy of 2 GeV/nucl..
Whereas usual QMD and RQMD calculations are assigned with QMD
and RQMD, respectively, QMD calculations with Lorentz contracted
starting
conditions (see text) are assigned with QMD*. RQMD
calculations with equal time constraints (see eq.(23) and text) are
assigned with RQMD*. The impact parameter was 2 fm and a hard EOS
was used.\\ \\
\end{document}